\documentclass[aps,floatfix,eqsecnum,showpacs,preprint,tightenlines]{revtex4}
\usepackage{epsfig}
\usepackage{amsmath}

\def\e{\kern+.6ex\lower.42ex\hbox{$\scriptstyle \iota$}\kern-1.20ex e}
\newcommand{\fet}[1]{\mbox{\boldmath $ #1 $}}

\newcommand{\NNNLO}{N$^3$LO~}
\newcommand{\NNLO}{N$^2$LO~}

\begin{document}

\title{
The triton with long-range chiral N$^{\bf 3}$LO three nucleon forces
}

\author{R.~Skibi\'nski$^1$}
\author{J.~Golak$^1$}
\author{K.~Topolnicki$^1$}
\author{H.~Wita{\l}a$^1$}

\author{E.~Epelbaum$^2$}
\author{W.~Gl\"ockle$^2$}
\author{H.~Krebs$^2$}

\author{A.~Nogga$^3$}

\author{H.~Kamada$^4$}

\affiliation{$^1$M. Smoluchowski Institute of Physics, Jagiellonian
University, PL-30059 Krak\'ow, Poland}

\affiliation{$^2$Institut f\"ur Theoretische Physik II,
Ruhr-Universit\"at Bochum, D-44780 Bochum, Germany}

\affiliation{$^3$Forschungszentrum J\"ulich,
          Institut f\"ur Kernphysik,
          Institute for Advanced Simulation
          and  J\"ulich Center for Hadron Physics, D-52425 J\"ulich, Germany}

\affiliation{$^4$ Department of Physics, Faculty of Engineering,
Kyushu Institute of Technology, Kitakyushu 804-8550, Japan}

\date{\today}

\begin{abstract}
Long-range contributions to the three-nucleon force that have been
recently worked out in chiral effective field theory at 
next-to-next-to-next-to-leading order are for the first time 
included in the triton and the doublet nucleon-deuteron
scattering length calculations. The strengths of the two short-range
terms available at this order in the chiral expansion are determined 
from the triton binding energy and the neutron-deuteron doublet
scattering length. The structure of the resulting three-nucleon force
is explored and effects for the two-nucleon correlation function in
the triton are investigated. Expectation values of the individual
contributions to the three-nucleon force in the triton are found to be
in the range from a few $100$ keV to about $1$ MeV. Our study
demonstrates that the
very complicated operator structure of the novel chiral
three-nucleon forces can be successively implemented in three-nucleon
Faddeev calculations.

\end{abstract}
\pacs{21.45.-v, 21.30.Fe, 21.45.Ff} \maketitle \setcounter{page}{1}

\section{Introduction}
Chiral effective field theory (EFT) provides a powerful framework to
systematically describe low-energy dynamics of few- and many-nucleon 
systems. Various variants of effective theories for nuclear forces 
have been explored,
see~\cite{Epelbaum,Epelbaum_review,Machleidt} for recent review
articles. Up to now, the most advanced few-nucleon studies have been carried out
within a framework based on pions and nucleons as the only explicit
degrees of freedom taken into account in the effective Lagrangian. 
Within this approach, the nucleon-nucleon (NN) force is currently
available up to next-to-next-to-next-to leading order
(\NNNLO)  in the chiral expansion. At this chiral
order, it receives contributions from one-, two- and three-pion
exchange diagrams as well as short-range NN contact interactions with
up to four derivatives. As demonstrated in
Refs.~\cite{NN_in_N3LO,Entem:2003ft},  NN phase
shifts are accurately described at \NNNLO up to laboratory energies of
the order of 200 MeV. 
The theoretical uncertainty due to truncation of the chiral expansion
is estimated in Ref.~\cite{NN_in_N3LO} by means of a cutoff
variation. Within the spectral function regularization (SFR) framework \cite{SFR}
adopted in Ref.~\cite{NN_in_N3LO}, the NN potential depends on two ultraviolet
cutoffs $\tilde \Lambda$ and $\Lambda$. 
The first one removes large-mass components in the spectrum
of the two-pion exchange potential which cannot be correctly described
within the chiral EFT framework while the other one provides
regularization of the Lippmann-Schwinger equation. 
Five different combinations of these cut-off
parameters are available for the NN potentials of
Ref.~\cite{NN_in_N3LO}. The residual dependence of low-energy observables on the
cutoff choice provides a measure of the importance of higher-order
contact interactions and thus may serve as an estimate of the
theoretical uncertainty.  

Parallel to these developments three-nucleon force (3NF) has
also been explored within the framework of chiral effective field theory. 
The first nonvanishing contributions to the 3NF emerge at
next-to-next-to-leading order (\NNLO)~\cite{3NF_in_N2LO} 
from 
the two-pion exchange and one-pion-exchange-contact diagrams as well
as the purely short-range derivative-less three-nucleon contact
interaction~\cite{3NF_in_N2LO}, see also Ref.~\cite{vanKolck:1994yi} for a
pioneering work along this line.  The resulting  \NNLO three-nucleon
potential depends  on two low-energy constants (LECs) $D$ and $E$ accompanying the
short-range $\pi NN$ and $NNN$ vertices, respectively.  The values
of these LECs need to be fixed from a fit to few-nucleon data. Among a
few possible observables 
that have been
used in this connection are the triton
binding energy and the nucleon-deuteron doublet scattering length
$^2a_{nd}$ \cite{3NF_in_N2LO,Epelbaum:2009zsa}, $\alpha$-particle binding energy
\cite{Nogga:2005hp,Epelbaum:2011md},  the properties of light nuclei
\cite{Navratil:2007we} and the triton beta decay \cite{Gazit:2008ma}.
The \NNLO 3NF of~\cite{3NF_in_N2LO} was successfully used in
three-body calculations, see Refs.~\cite{Kistryn,Kalantar} for a few
examples of recent studies. At this order, chiral EFT yields a good
description of elastic scattering and deuteron breakup observables up
to energies of about $\approx
50$ MeV. The accuracy of the results in this regime is comparable with
the one that is achieved by realistic phenomenological NN and 3N
interactions such as e.g.~AV18~\cite{AV18} 2NF in combination with UrbanaIX~\cite{Urbana} 3NF 
or CD-Bonn~\cite{CDBonn} 2NF in combination with the Tucson-Melbourne~\cite{TM} 3NF, see 
\cite{Witala.2001,Kistryn}.
However, the spread of the results is relatively large for some spin
observables
which clearly calls for the inclusion of new terms of the nuclear
interaction that occur at higher orders of the chiral expansion.

Subleading contributions to the 3NF are currently being investigated
by several groups. At \NNNLO, one has to take into account 
(irreducible) contributions emerging from all possible one-loop
three-nucleon diagrams constructed with the lowest order vertices. In
addition, there are (tree-level) leading relativistic corrections, 
see \cite{Friar:1994zz} for an early work on the longest-range relativistic
corrections. Note that the tree diagrams involving higher-order
vertices from the effective chiral Lagrangian do not produce any
irreducible pieces. Effects due to two-pion exchange 3NF in elastic
nucleon-deuteron scattering were already explored by Ishikawa and
Robilotta  \cite{Ishikawa} within a hybrid approach and found to be
rather small. 
The \NNNLO contributions feed into five different topologies as will
be explained in detail in the next section. The
explicit expressions both in momentum and in coordinate space for the
long-range contributions have already been worked out \cite{Ishikawa,Bernard}. 
Their inclusion in numerical few-body calculations appears to be
challenging due to the very rich and complicated operator structure. 
The large number of terms in the 3NF at  \NNNLO, see Ref.~\cite{Bernard},
requires an efficient method of performing the partial wave
decomposition. Recently such a method has been proposed
\cite{Golak.2010} and tested for the Tucson-Melbourne
force~\cite{aPWDTM}. Here and in what follows, this approach will be
referred to as the automatized
partial wave decomposition (aPWD). In this paper we
apply this method of the numerical partial wave decomposition to the
\NNNLO 3NF contributions derived in ~\cite{Bernard}. For
the first time, the parts of 3NF at \NNNLO different from the two-pion exchange
force are included in the triton and the scattering length
calculations. In order to test the implementations and get a 
first hint to possible effects of these forces, 
we  fix the two LECs entering the 3NF from the triton
binding energy and the nucleon-deuteron doublet scattering length and
explore the effects due to these novel 3NF terms by computing
the $^3$H properties. Although this calculations is still incomplete
since not all 3NF contributions at \NNNLO are taken into account, it
provides an important first step towards the complete \NNNLO analysis of 3N
scattering and demonstrates our ability to numerically handle the
rather complicated structure of the subleading chiral 3NF. 

Our paper is organized as follows. In Sect.~\ref{section1} we
describe briefly the structure of the chiral 3NF at \NNNLO. In
Sect.~\ref{section3} we discuss in detail the 
partial wave decomposition needed in our scattering and
bound-state calculations. Next, the procedure of fixing the LECs is 
described in Sect.~\ref{section4} where also the obtained values of LECs are
listed. These results  are used in Sect.~\ref{section5} to explore some
properties of the triton. Finally, our
findings are  summarized in Sec.~\ref{conclusion}.

\section{3NF at N$\fet{^3}$LO }
\label{section1}

The subleading (i.e.~\NNNLO) contributions to the 
three-nucleon force $V_{123}$ can be written in the form~\cite{Bernard}
\begin{equation}
V_{123}=V_{2\pi}+V_{2\pi-1\pi}+V_{ring}+V_{1\pi-cont}+V_{2\pi-cont}+
V_{1/m}\;.
\end{equation}
where the individual terms refer, in order, to the  two-pion
exchange, two-pion-one-pion-exchange, ring (i.e.~one pion being exchanged
between each of the three nucleon pairs), one-pion-exchange-contact
and two-pion-exchange-contact contributions as well as the leading
relativistic corrections, see  Fig.~1 of Ref.~\cite{Bernard} for a
diagrammatic representation.   
The expressions for the (static) long-range part of the 3NF given by
the first three terms in the above equation have been worked out in
heavy-baryon chiral perturbation theory in Ref.\cite{Bernard}. The
two-pion exchange contribution at the one-loop level has also been
calculated within the infrared-regularized version of chiral
perturbation theory in Ref.~\cite{Ishikawa}. The shorter-range
contributions involving two-nucleon contact interactions 
and relativistic corrections 
are currently being worked out \cite{Evgeny_private}.

While the two-pion-exchange $V_{2\pi}$  and one-pion-exchange-contact parts 
$V_{1\pi-cont}$ already occur at \NNLO and receive corrections at
\NNNLO, the remaining topologies first emerge at \NNNLO. 
It is important to emphasize that all subleading contributions to the
3NF are parameter-free. Thus, the low-energy constants $D$ and $E$
entering the one-pion-exchange-contact and the purely short-range
parts of the 3NF at \NNLO are the only unknown parameters up to \NNNLO.   

Here and in what follows, we adopt the notation in which a given 3NF $V_{123}$ 
is decomposed into three terms 
\begin{eqnarray}
V_{123} = V^{(1)} +  V^{(2)} +  V^{(3)} ~, \label{e3nf_split}
\end{eqnarray}
where each $V^{(i)}$ is symmetrical under interchanging the nucleons 
$j$ and $k$ ($i, j, k = 1, 2, 3$, $i \ne j \ne k$). Clearly, this condition
does not specify $V^{(1)}$ uniquely. In the following we choose
$V^{(1)}$ in such a way that the number of operator structures is
minimized which is convenient for the aPWD.

The operator structure of the 2$\pi$ exchange part $V^{(1)}_{2\pi}$
at \NNNLO  remains the same as at \NNLO
\begin{equation}
V^{(1)}_{2\pi}=F_1 \; \vec{\sigma_2}\cdot\vec{q_2} \;
\vec{\sigma_3} \cdot\vec{q_3} \; \fet{\tau_2} \cdot \fet{\tau_3} 
+ F_2 \; \vec{\sigma_2}\cdot\vec{q_2} \;
\vec{\sigma_3} \cdot\vec{q_3} \; \vec{q_2} \times \vec{q_3} \cdot \vec{\sigma_1} \; \fet{\tau_2} \times \fet{\tau_3} \cdot \fet{\tau_1},
\end{equation}
where $\vec{q_i}$ is the momentum transfer to the i-th nucleon,
$\vec{q_1}+\vec{q_2}+\vec{q_3}=0$ and $\vec{\sigma_i}$
($\fet{\tau_i}$) are Pauli spin (isospin) matrices for nucleon $i$.
The scalar functions $F_1=F_1(q_2,q_3,\hat{q_2}\cdot\hat{q_3})$
and $F_2=F_2(q_2,q_3,\hat{q_2}\cdot\hat{q_3})$ depend on
the  LECs  $\tilde{c}_{1,3,4}$ which accompany the subleading pion-nucleon
vertices. Chiral expansion of $F_1$ and $F_2$ up to \NNNLO  
has the form~\cite{Bernard} 
\begin{eqnarray}
F_1 &=& \frac{g_A^4}{4 F_{\pi}^4} \;\frac{(-4\tilde{c_1} M_{\pi}^2 + 2\tilde{c_3}\vec{q_2}\cdot\vec{q_3})}
{(q_2^2+M_{\pi}^2)(q_3^2+M_{\pi}^2)} + \tilde{F_1} \label{eqF1F21} \\
F_2 &=& \frac{g_A^4}{4 F_{\pi}^4}
\;\frac{\tilde{c_4}}{(q_2^2+M_{\pi}^2)(q_3^2+M_{\pi}^2)} +
\tilde{F_2} \;, \label{eqF1F22}
\end{eqnarray}
with
\begin{eqnarray}
\tilde{F_1} &=& \frac{g_A^4}{128 \pi F_{\pi}^6} \frac{1}{(q_2^2+M_{\pi}^2)(q_3^2+M_{\pi}^2)}
(M_{\pi}(M_{\pi}^2+3q_2^2+3q_3^2+4\vec{q_2}\cdot\vec{q_3}) \nonumber \\
&+& (2M_{\pi}^2+q_2^2+q_3^2+2\vec{q_2}\cdot\vec{q_3})(3M_{\pi}^2+3q_2^2+3q_3^2+4\vec{q_2}\cdot\vec{q_3})
A(\vert \vec{q_2}+\vec{q_3} \vert)) \\
\tilde{F_2} &=& \frac{-g_A^4}{128 \pi F_{\pi}^6} \frac{1}{(q_2^2+M_{\pi}^2)(q_3^2+M_{\pi}^2)}
(M_{\pi}+(4M_{\pi}^2+q_2^2+q_3^2+2\vec{q_2}\cdot\vec{q_3})A(\vert \vec{q_2}+\vec{q_3} \vert)) \,,
\end{eqnarray}
where the loop function $A(q)$ is defined as 
\begin{eqnarray}
A(q) &=& \frac{1}{2q} \arctan{\frac{q}{2M_{\pi}}} \,.
\end{eqnarray}
The axial-vector coupling
constant, the weak pion decay constant and the pion mass are denoted as
$g_A$, $F_{\pi}$ and $M_{\pi}$, respectively. 
Note that SFR changes $A(q)$ (see~\cite{SFR}). For the study here we do not
need to consider this change since they differ only by higher order
polynomials~\cite{Epelbaum}.
The quantities $\tilde
c_i$ appearing in the above expressions are related to the \NNLO
LECs $c_i$ entering the effective chiral Lagrangian via    
\begin{eqnarray}
\label{ci_def}
\tilde{c_1} &=& c_1-\frac{g_A^2 M_{\pi} }{64 \pi F_{\pi}^2} = -0.94\; {\rm GeV}^{-1} \nonumber \\
\tilde{c_3} &=& c_3+\frac{g_A^4 M_{\pi} }{16 \pi F_{\pi}^2} = -2.51\; {\rm GeV}^{-1}  \nonumber \\
\tilde{c_4} &=& c_4-\frac{g_A^4 M_{\pi} }{16 \pi F_{\pi}^2} =
2.51\; {\rm GeV}^{-1} \;, \nonumber
\end{eqnarray}
where we have adopted the values for the $c_i$ from
Ref.~\cite{NN_in_N3LO}, namely
\begin{equation}
c_1 = -0.81\; \mbox{GeV}^{-1}, \quad \quad
c_3 = -3.40\; \mbox{GeV}^{-1}, \quad \quad
c_4 = 3.40\; \mbox{GeV}^{-1}\,.
\end{equation}
These values are consistent with the ones determined from pion-nucleon
scattering \cite{Buettiker:1999ap}. The finite shifts of the LECs
$c_i$ in Eq.~(\ref{ci_def}) emerge from pion loops at \NNNLO. 
  
The $V^{(1)}_{2\pi-1\pi}$ interaction at \NNNLO  has the following
operator structure
\begin{eqnarray}
V^{(1)}_{2\pi-1\pi}&=&
 \frac{\vec \sigma_1 \cdot \vec q_1}{q_1^2 + M_\pi^2} \Big[ \fet \tau_2
  \cdot \fet \tau_1  \; \left( \vec \sigma_3 \cdot \vec q_2 \; \vec q_2 \cdot
    \vec q_1 \; F_1 (q_2)  + \vec \sigma_3 \cdot \vec q_2 \; F_2 (q_2) +
  \vec \sigma_3 \cdot \vec q_1 \;  F_3 (q_2)  \right) \nonumber \\
 &+& \fet \tau_3
  \cdot \fet \tau_1 \; (  \vec \sigma_2 \cdot \vec q_2 \;  \vec q_2 \cdot
    \vec q_1 \; F_4 (q_2) 
+  \vec \sigma_2 \cdot \vec q_1 \; F_5  (q_2) +
\vec \sigma_3 \cdot \vec q_2 \;  F_6 (q_2) +
  \vec \sigma_3 \cdot \vec q_1 \;   F_{7} (q_2)
)  \nonumber \\
&+& \fet \tau_2
  \times \fet \tau_3 \cdot \fet \tau_1 \; \vec \sigma_2 \times \vec \sigma_3
  \cdot \vec q_2 \;  F_{8} (q_2)
\Big] + (2 \leftrightarrow 3) \,.
\end{eqnarray}
with scalar functions $F_i(q_i)$. The interchange of nucleons $(2
\leftrightarrow 3)$ refers to the interchange of momentum vectors,
spin and isospin matrices and arguments of the functions $F_i$. 
Explicit expressions for the scalar functions $F_i(q_i)$ which appear in 
the $V^{(1)}_{2\pi-1\pi}$ can be found
in Ref.~\cite{Bernard}.

The $V^{(1)}_{ring}$ force is chosen as
\begin{eqnarray}
V^{(1)}_{ring} &=&
\vec{\sigma}_1\cdot\vec{\sigma}_2 \; {\fet \tau}_2\cdot{\fet\tau}_3  \; R_1+
\vec{\sigma}_1\cdot\vec{q}_1\vec{\sigma}_2\cdot\vec{q}_1 \; {\fet
  \tau}_2\cdot{\fet\tau}_3  \; R_2+
\vec{\sigma}_1\cdot\vec{q}_1\vec{\sigma}_2\cdot\vec{q}_3 \; {\fet
  \tau}_2\cdot{\fet\tau}_3  \; R_3 \nonumber \\
&+&
\vec{\sigma}_1\cdot\vec{q}_3\vec{\sigma}_2\cdot\vec{q}_1 \; {\fet
  \tau}_2\cdot{\fet\tau}_3  \; R_4
+\vec{\sigma}_1\cdot\vec{q}_3\vec{\sigma}_2\cdot\vec{q}_3 \; {\fet
  \tau}_2\cdot{\fet\tau}_3  \; R_5+{\fet\tau}_1\cdot{\fet\tau}_3 \;  R_6
+\vec{\sigma}_1\cdot\vec{q}_1\vec{\sigma}_3\cdot\vec{q}_1  \; R_7 \nonumber \\
&+& \vec{\sigma}_1\cdot\vec{q}_1\vec{\sigma}_3\cdot\vec{q}_3  \; R_8
+\vec{\sigma}_1\cdot\vec{q}_3\vec{\sigma}_3\cdot\vec{q}_1  \; R_9
+\vec{\sigma}_1\cdot\vec{\sigma}_3  \; R_{10}
+\vec{q}_1\cdot \vec{q}_3\times\vec{\sigma}_2 \;
{\fet\tau}_1\cdot{\fet\tau}_2\times{\fet\tau}_3 \;  R_{11} \nonumber \\
&+&{\fet \tau}_1\cdot{\fet\tau}_2  \; S_1+
\vec{\sigma}_1\cdot\vec{q}_1\vec{\sigma}_3\cdot\vec{q}_1 \; {\fet\tau}_1\cdot{\fet\tau}_2
 \; S_2+
\vec{\sigma}_1\cdot\vec{q}_3\vec{\sigma}_3\cdot\vec{q}_1 \; {\fet\tau}_1\cdot{\fet\tau}_2
 \; S_3+
\vec{\sigma}_1\cdot\vec{q}_1\vec{\sigma}_3\cdot\vec{q}_3 \; {\fet\tau}_1\cdot{\fet\tau}_2
 \; S_4\nonumber \\
&+&
\vec{\sigma}_1\cdot\vec{q}_3\vec{\sigma}_3\cdot\vec{q}_3 \; {\fet\tau}_1\cdot{\fet\tau}_2
 \; S_5+
\vec{\sigma}_1\cdot\vec{\sigma}_3 \; {\fet\tau}_1\cdot{\fet\tau}_2
 \; S_6+
\vec{q}_1\cdot \vec{q}_3\times\vec{\sigma}_1  \;
{\fet\tau}_1\cdot {\fet\tau}_2\times{\fet\tau}_3  \;  S_7 \nonumber \\
&+& (2 \leftrightarrow 3)\,,
\end{eqnarray}
where the expressions for the scalar functions
$R_i=R_i(q_1,q_3,\hat{q_1} \cdot \hat{q_3})$ and
$S_i=S_i(q_1,q_3,\hat{q_1} \cdot \hat{q_3})$ can be found in
Ref.~\cite{Bernard}.

The modifications of the one-pion-exchange-contact
$V^{(1)}_{1\pi-cont}$ term arising at \NNNLO  are in preparation
~\cite{Evgeny_private}. Thus, instead of the full
$V^{(1)}_{1\pi-cont}$ interaction we use the lowest-order result for
it resulting at 
\NNLO\cite{3NF_in_N2LO}:
\begin{equation}
V^{(1)}_{d-term} = - \frac{g_A \, D}{8 F_\pi^2}\;
\frac{\vec \sigma_1\cdot \vec q_1}{q_1^2 + M_\pi^2} \;
 (\fet \tau_1 \cdot \fet \tau_3 \; \vec \sigma_3 \cdot \vec q_1 +
         \fet \tau_1 \cdot \fet \tau_2 \; \vec \sigma_2 \cdot \vec
         q_1)\;.
\end{equation}
The low energy constant $D$ can be expressed as
$D=c_D/(F_{\pi}^2 \Lambda_{\chi})$, where $c_D$ is a
dimensionless free parameter and the chiral symmetry breaking scale
$\Lambda_{\chi}$ is estimated to be $\Lambda_{\chi}=700$ MeV. Here and
in what follows, we use $F_{\pi}=92.4$ MeV for the pion decay constant. 
The value of $c_D$ has to be determined
from experimental data. This is described in section \ref{section4}
for our fit for the test case of this study. Of course, results of this 
fit will significantly change when  the complete short range interaction is taken into account.

As already pointed out above, the remaining terms $V^{(1)}_{2\pi-cont}$ and
$V^{(1)}_{1/m}$ are also not available yet
\cite{Evgeny_private} and thus cannot be taken into account in the
present study. Finally, the purely short-range part of the 3NF has the
form 
\cite{3NF_in_N2LO}:
\begin{eqnarray}
V^{(1)}_{e-term} &=& E \fet \tau_2 \cdot \fet \tau_3 \;.
\end{eqnarray}
Again, the LEC  $E$ is usually expressed in terms of a
dimension-less parameter $c_E$ via $E=c_E/(F_{\pi}^4
\Lambda_{\chi})$ which needs to be determined from (at least)
three-nucleon data. 

To summarize, in this paper we
use the $V^{(1)}_{2\pi}$, $V^{(1)}_{2\pi-1\pi}$, $V^{(1)}_{ring}$ \NNNLO 
terms combined with the $V^{(1)}_{d-term}$ and $V^{(1)}_{e-term}$ terms
at N$^2$LO. The \NNNLO contributions to 
$V^{(1)}_{1\pi-cont}$ and $V^{(1)}_{2\pi-cont}$ and relativistic
corrections $V^{(1)}_{1/m}$ are not included. Finally, the remaining 
terms $V^{(2)}$ and $V^{(3)}$  of
Eq.~(\ref{e3nf_split}) can be obtained from $V^{(1)}$ by appropriate
permutations of nucleons.

\section{Numerical calculations of 3NF matrix elements}
\label{section3}

 We work in the momentum space using three-nucleon
partial-wave states $\mid p, q, \alpha \rangle$ in the
$jJ$-coupling~ \cite{book,physrep_96}
\begin{eqnarray}
\mid p, q, \alpha \rangle \equiv
\mid  p q (l s ) j (\lambda \frac12 ) I (j I ) J M_J \rangle \mid (t \frac12 )
T M_T \rangle  \ ,
\label{eqn.alpha}
\end{eqnarray}
where $p$ and $q$ are magnitudes of the standard Jacobi momenta and
$\alpha$ denotes a set of discrete quantum numbers defined in the following way:
the spin $s$ of the subsystem composed from nucleons 2 and 3
is coupled with their orbital angular momentum $l$
to the total angular momentum $j$. The spin $1/2$
of the spectator particle $1$
couples with its relative orbital angular momentum $\lambda$
to the total angular momentum $I$ of nucleon $1$.
Finally, $j$ and $I$ are coupled to the total 3N
angular momentum $J$ with the projection $M_J$. For the isospin part,
 the total isospin $t$ of the  subsystem $(23)$ is coupled with
the isospin $1/2$ of
the spectator nucleon to the total 3N isospin $T$ with the projection $M_T$.

The matrix elements of $V^{(1)}_{2\pi},V^{(1)}_{2\pi-1\pi}$ and
$V^{(1)}_{ring}$ forces in the basis $\mid p, q, \alpha \rangle$ are
obtained using the recently proposed aPWD method 
\cite{Golak.2010,aPWDTM}. In this approach the spin-momentum and
isospin parts of three-nucleon interactions are calculated using a
software for symbolic calculations. The resulting momentum-dependent
functions are then integrated numerically in five dimensions over
angular variables. The major advantage of this method is its
generality. It can be applied to \emph{any} momentum-spin-isospin
operator including, in particular, the full operator structure of the
3NF at \NNNLO including even the non-local relativisitic corrections. 
The only complication emerges in the treatment of
the ring contributions to the 3NF due to rather complex
expressions for  the functions $R_i=R_i(q_1,q_3,\hat{q_1} \cdot \hat{q_3})$ and
$S_i=S_i(q_1,q_3,\hat{q_1} \cdot \hat{q_3})$ which involve certain
scalar integrals related to the three-point function. For any given
values of the arguments, these integrals have to be computed
numerically. This evaluation is too expensive to be carried out on-the-fly
 during the aPWD. Moreover, while  the
functions $R_i$ and $S_i$  are, of course, finite and smooth for all
possible values of 
their arguments, they are given in Ref.~\cite{Bernard} as linear
combinations of terms, some of which are becoming singular under
certain kinematical conditions. Their numerical implementation
therefore requires special care.  In order to deal with these
difficulties, we first evaluate the functions  $R_i$ and $S_i$ at a
(fixed) dense grid of points for their arguments and then use
interpolation to compute them for arbitrary values of $q_1$,
$q_3$ and $\hat{q_1}\cdot \hat{q_3}$ as needed in the aPWD approach. We have
carefully checked the stability of this procedure by increasing the
density of the grid points. 
Finally, the partial wave decomposition of $V^{(1)}_{d-term}$ and
$V^{(1)}_{e-term}$ is performed with the standard techniques
~\cite{3NF_in_N2LO} but also verified with the new method.

\begin{figure}[t]\centering
\epsfig{file=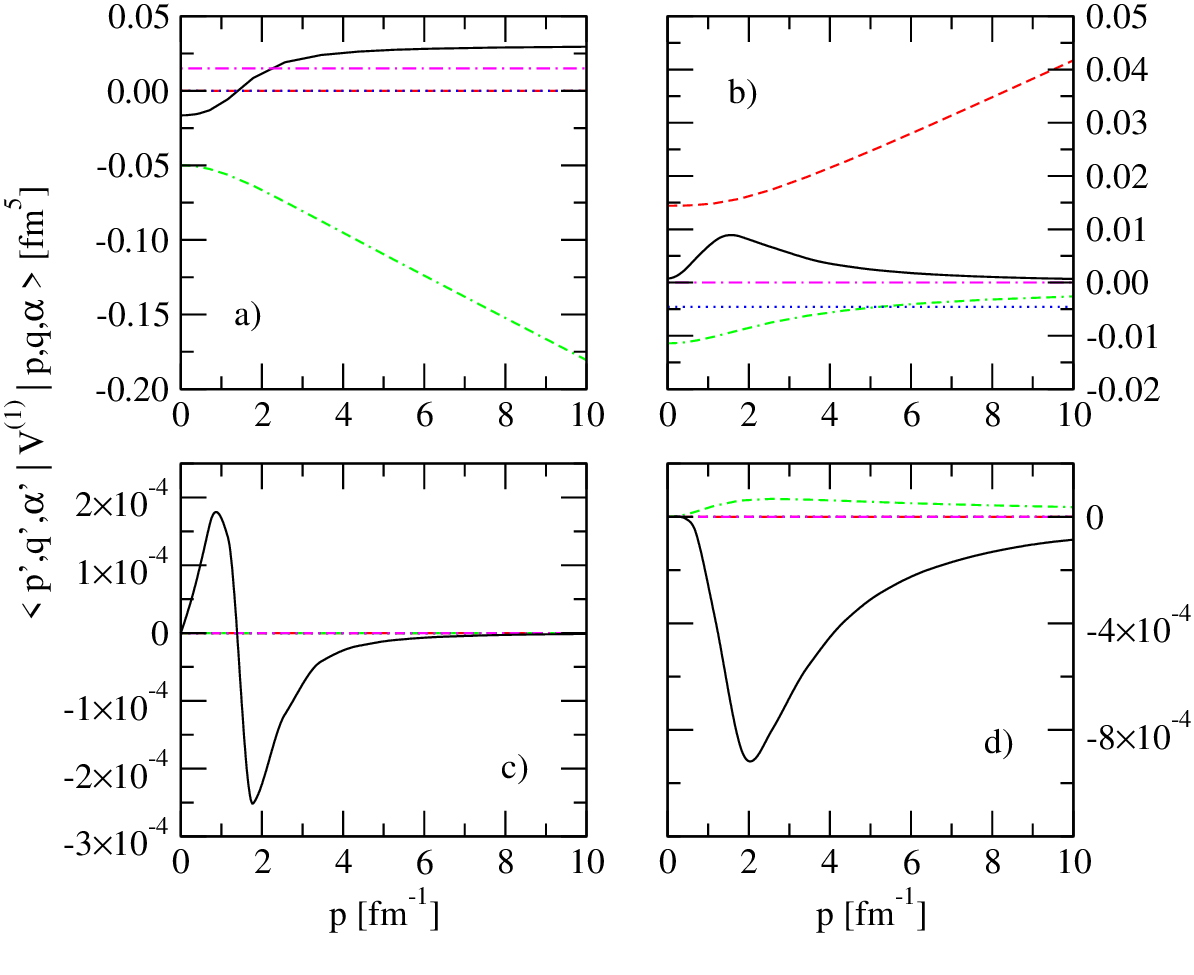,width=15cm,clip=true}
\caption{(Color online) The \NNNLO  3NF matrix elements (before regularization) 
\mbox{$< p',q',\alpha' \vert V^{(1)} \vert p, q,\alpha >$} 
as a function of the momentum $p$
for 
$p'=0.268$ fm$^{-1}, q'=2.842$ fm$^{-1}$ and $q=0.132$ fm$^{-1}$ 
and different combinations ($\alpha', \alpha$): a) (1,1), b) (1,5), c)
(4,4) and  d) (4,7). The solid (black), dashed(red),
double-dash-dotted(green), dotted(blue) and dash-dotted (magenta)
 line shows the $V_{2\pi}$,$V_{2\pi-1\pi}$,$V_{ring}$,$V_{d-term}$ and
 $V_{e-term}$ components, respectively. 
}
\label{fig1.1}
\end{figure}

Examples of the resulting matrix elements $\langle p',q',\alpha'
\vert V_i^{(1)} \vert p,q,\alpha \rangle$ are given in
Fig.~\ref{fig1.1} as a function of the momentum $p$. Here, we fix  the
momenta to be $p'=0.268$ fm$^{-1}$, $q'=2.842$ fm$^{-1}$ and $q=0.132$
fm$^{-1}$ and consider the following four channel combinations: ($\alpha'=\alpha=1$),
($\alpha'=1,\alpha=5$), ($\alpha'=\alpha=4$) and
($\alpha'=4,\alpha=7$). These channels correspond to the quantum
numbers given in Tab.~\ref{tab0}.
\begin{table}
\begin{tabular}{|c|c|c|c|c|c|c|}
\hline
$\;\;\;\;\alpha$\;\;\;\; & \;\;\;\;l\;\;\;\; & \;\;\;\;s\;\;\;\; & \;\;\;\;j\;\;\;\; &
 \;\;\;\;$\lambda$\;\;\;\; & \;\;\;\;I\;\;\;\; & \;\;\;\;t\;\;\;\; \\
\hline
1 & 0 & 0 & 0 & 0 & $\frac12$ & 1 \\
4 & 1 & 0 & 1 & 1 & $\frac32$ & 0 \\
5 & 0 & 1 & 1 & 0 & $\frac12$ & 0 \\
7 & 2 & 1 & 1 & 0 & $\frac12$ & 0 \\
\hline
\end{tabular}
\caption{The values of the discrete quantum numbers for the selected
$\alpha$ states (\ref{eqn.alpha})  with the total angular momentum
$J=1/2$, the total isospin $T=1/2$ and its projection
$M_T=-1/2$.}
\label{tab0}
\end{table}
%
Channels $\alpha=1$ and $\alpha=5$ are especially important since
those two states provide the dominant components of the $^3$H wave function.
As can be noticed, all three components of the \NNNLO  3NF give a strong
contribution: the $V^{(1)}_{ring}$ dominates for the channel
combination ($\alpha'=\alpha=1$), $V^{(1)}_{2\pi-1\pi}$ for
($\alpha'=1,\alpha=5$) and $V^{(1)}_{2\pi}$ for ($\alpha'=\alpha=4$)
and ($\alpha'=4,\alpha=7$). 
We emphasize, however, that the large size of these matrix elements
(which contain certain admixtures of short-range operators) as
compared to the N$^2$LO terms does not necessarily imply that theirs effects in
low-energy observables are large.  The values of the LECs for
$V^{(1)}_{d-term}$ and $V^{(1)}_{e-term}$ terms shown in
Fig.~\ref{fig1.1} are set to be $c_D=1$ and $c_E=1$ in order to allow
for a qualitative comparison of the strength of the individual terms. Their real
contributions emerging after fitting the LECs $c_D$ and $c_E$ to
experimental data will be discussed in the next section.
Fig.~\ref{fig1.1} also clearly demonstrates that not all terms
contribute to each channel combination due to the spin-isospin
dependence. The $V^{(1)}_{2\pi-1\pi}$ and $V^{(1)}_{d-term}$ terms contribute to $V^{(1)}$ only to
($\alpha'=1,\alpha=5$) and the $V^{(1)}_{e-term}$ term only
contributes to 
($\alpha'=\alpha=1$). Further, $V^{(1)}_{ring}$ vanishes for $\alpha'=\alpha=4$.

\begin{figure}[t]\centering
\epsfig{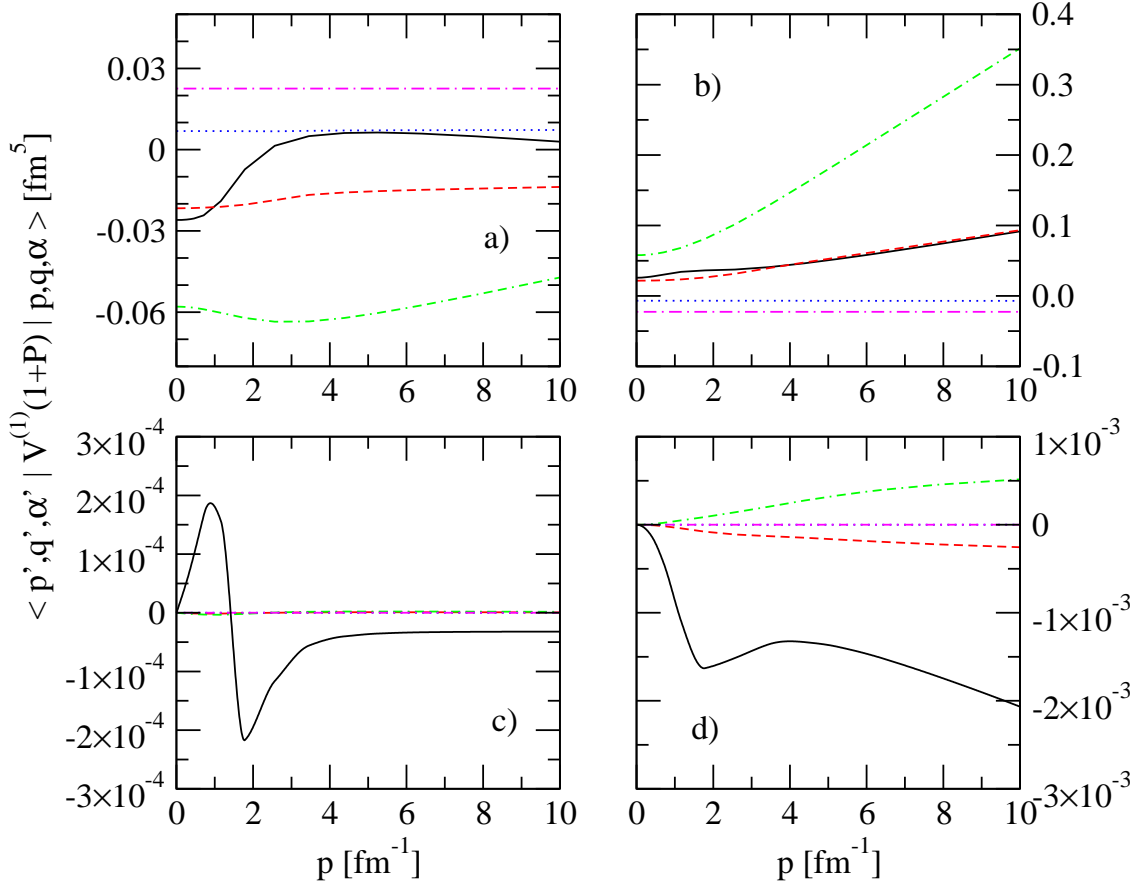}
\caption{(Color online) The \NNNLO  3NF matrix elements (before regularization)
\mbox{$< p',q',\alpha' \vert V^{(1)}(1+P) \vert p, q,\alpha >$}
as a function of the momentum $p$. The momenta $p',q'$ and $q$, 
channel combinations ($\alpha', \alpha$) and lines are the same as 
in Fig.~\ref{fig1.1}.
}
\label{fig1.1pp}
\end{figure}

The component $V^{(1)}$ of $V_{123}$ enters the dynamical equations
for 3N bound and scattering states~\cite{physrep_96, Nogga.1997}
(see below) only in combination with the permutation operator $P$ forming
the operator $V^{(1)}(1+P)$. The permutation operator $P \equiv
P_{12}P_{23} + P_{13}P_{23}$ is built from the transpositions
$P_{ij}$, which interchange nucleons $i$ and $j$.
The aPWD scheme can be used to obtain directly the $V^{(1)}(1+P)$ matrix
elements~\cite{aPWDTM} which allows us to avoid uncertainties
associated with the partial wave decomposition of the permutation operator. The
resulting matrix elements $\langle p',q',\alpha' \vert V^{(1)}(1+P)
\vert p,q,\alpha \rangle$ are given in Fig.~\ref{fig1.1pp} for the
same momenta and channel combinations as in Fig.~\ref{fig1.1}.
Again, we observe that matrix elements of all contributions to the 3NF
are large for these momenta. The
$V^{(1)}_{ring}$ contribution is particularly large in matrix elements with
$\alpha'=1$. For these channels also $V^{(1)}_{2\pi}$ and
$V^{(1)}_{2\pi-1\pi}$ are significant. Moreover for
$\alpha'=\alpha=1$ also $V^{(1)}_{d-term}$ and $V^{(1)}_{e-term}$
are non-negligible, at least with $c_D=1$ and $c_E=1$. For
$\alpha'=4$, the $V^{(1)}_{2\pi}$ piece dominates. Nevertheless in the case
of $\alpha=7$ also $V^{(1)}_{ring}$ provides a significant
contribution. Due to their spin-isospin structure the 
$V^{(1)}_{d-term}$ and $V^{(1)}_{e-term}$ forces are absent for 
($\alpha'=\alpha=4$) and ($\alpha'=4,\alpha=7$) channel combinations.

\begin{figure}[t]\centering
\epsfig{file=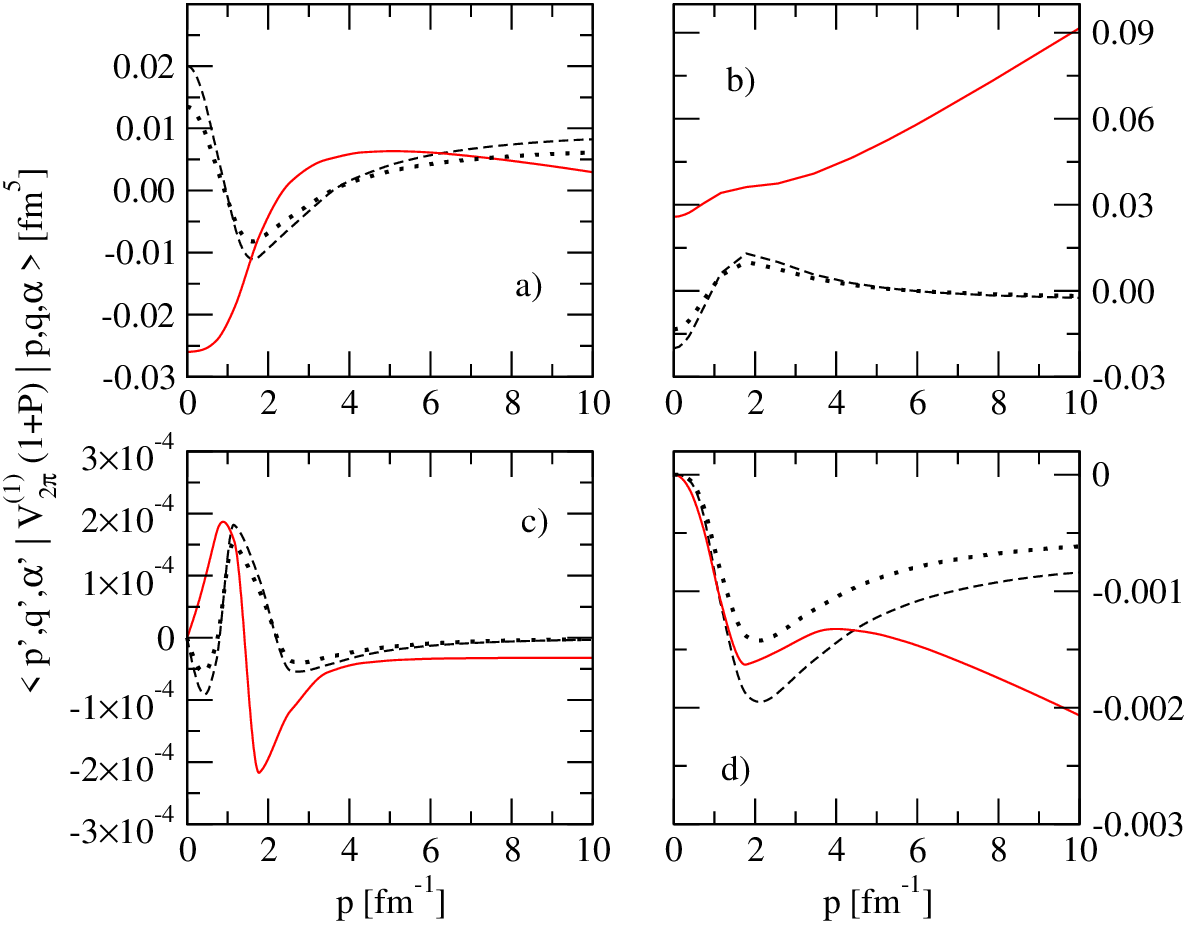,width=15cm,clip=true}
\caption{(Color online) The 2$\pi$-exchange part of 3NF matrix
elements (before regularization)
\mbox{$< p',q',\alpha' \vert V^{(1)}_{2\pi}(1+P) \vert p, q,\alpha >$}
as a function of the $p$ momentum. The momenta $p',q',q$ and
channel combinations ($\alpha', \alpha$)
 are the same as in Fig.~\ref{fig1.1}.
The dashed (black) and solid (red)
lines represent \NNLO and \NNNLO  results, respectively. The dotted
(black) line describes the matrix elements of $V^{(1)}_{2\pi}(1+P)$
obtained with $\tilde{F_1}$ and $\tilde{F_2}$ set artificially to
zero (see text). } \label{fig6.1pp.2pi}
\end{figure}

The 2$\pi$-exchange contribution at \NNNLO  can be compared with a
corresponding part of the \NNLO interaction. The example is given
in Fig.~\ref{fig6.1pp.2pi}, where the dashed (solid) line represents
the predictions obtained at \NNLO(\NNNLO). The $V^{(1)}_{2\pi}(1+P)$
matrix elements differ significantly for all considered channel
pairs. Fig.~\ref{fig6.1pp.2pi} demonstrates  that
these differences mainly originate from the parts of the
$V^{(1)}_{2\pi}$ interaction proportional to the $\tilde{F_1}$
and $\tilde{F_2}$ formfactors, see
Eqs.(\ref{eqF1F21})-(\ref{eqF1F22}). In particular, in
Fig.~\ref{fig6.1pp.2pi}, we also show results for
$V^{(1)}_{2\pi}(1+P)$ at \NNNLO  but with $\tilde{F_1}$ and
$\tilde{F_2}$ being artificially set to zero. In such a case the only
difference between matrix elements at \NNLO and \NNNLO  comes from the
different values of $c_i$ and $\tilde{c_i}$ LECs. The presented
matrix elements have then a similar dependence on momentum $p$ as
well as a similar magnitude in \NNNLO and \NNLO. 
Notice, however, that these observations
do not necessarily imply that the N$^3$LO corrections to the 3NF lead
to large effects in low-energy three-nucleon observables. In fact, the
opposite was observed in  Ref.\cite{Ishikawa}  for the case of the
two-pion exchange topology.   

The $V^{(1)}_i$ and $V^{(1)}_i(1+P)$ matrix elements shown in
Figs.~\ref{fig1.1} and~\ref{fig1.1pp} have to be regularized prior to
being used as input to the dynamical equations~\cite{3NF_in_N2LO}. We use the
regulator function of the form~\cite{Epelbaum}
\begin{equation}
f(p,q)=\exp{\frac{-(4p^2+3q^2)^3}{(4\Lambda^2)^3}}
\end{equation}
which ensures that the large momenta are sufficiently suppressed. Following
Ref.~\cite{Epelbaum} we use three values of $\Lambda$ parameter:
$450$, $550$ and $600$ MeV. The regularization transforms matrix
elements as
\begin{equation}
\langle p',q',\alpha' \vert V^{(1)}(1+P) \vert p,q,\alpha \rangle \rightarrow
f(p',q') \langle p',q',\alpha' \vert V^{(1)}(1+P) \vert p,q,\alpha \rangle f(p,q).
\label{eq.reg}
\end{equation}

\begin{figure}[t]\centering
\epsfig{file=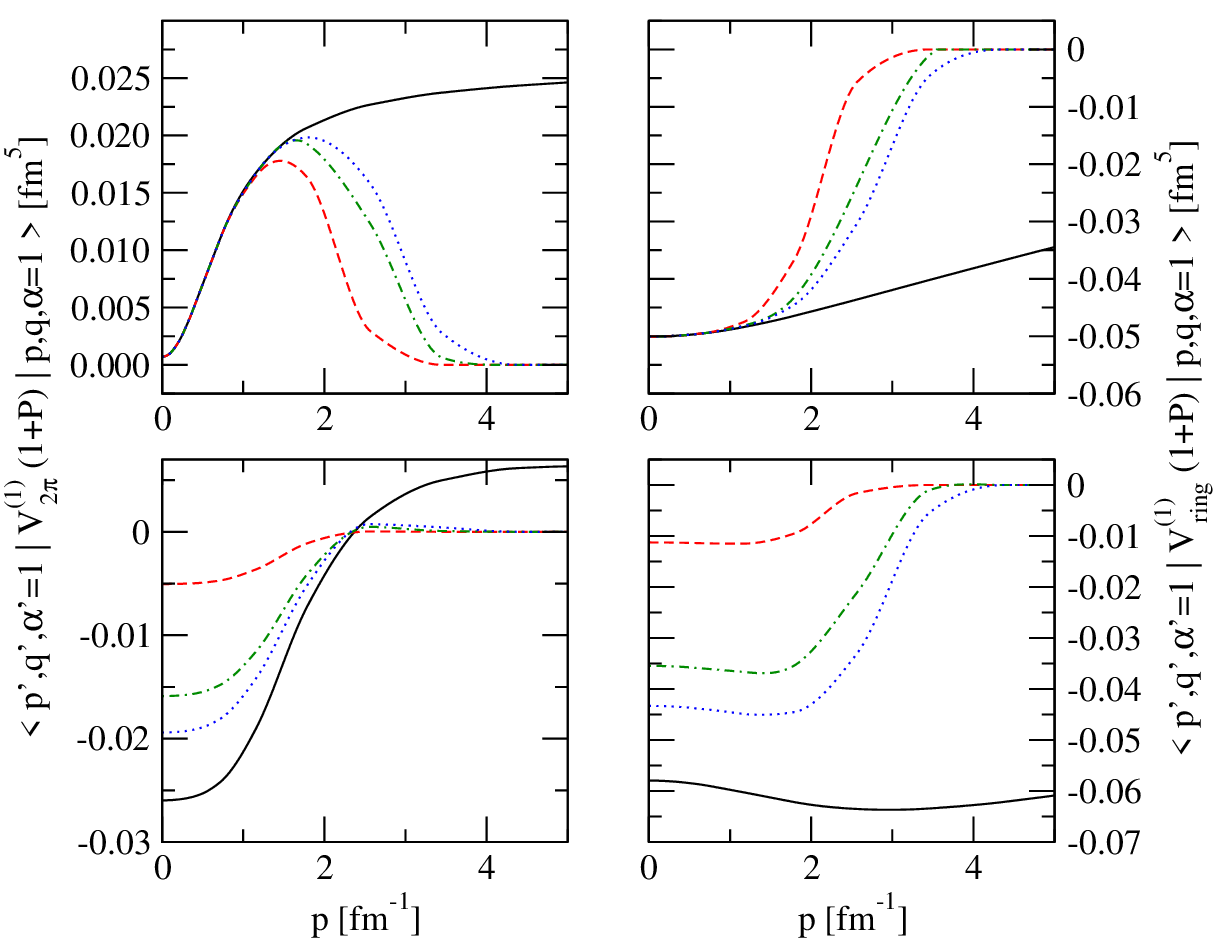,width=15cm,clip=true}
\caption{(Color on-line) The \NNNLO  3NF matrix elements
\mbox{$< p',q',\alpha'=1 \vert V^{(1)}_i(1+P) \vert p, q,\alpha=1 >$}
as a function of the $p$ momentum for the momenta 
$p'=q'=q=0.132$ fm$^{-1}$ (top) and
$p'=0.268$ fm$^{-1}, q'=2.842$ fm$^{-1}$ and $q=0.132$ fm$^{-1}$ (bottom).
The two components of the \NNNLO  3NF are shown: $V^{(1)}_{2\pi}(1+P)$ (left)
and $V^{(1)}_{ring}(1+P)$ (right).
The solid (black) line represents $V^{(1)}_{2\pi}(1+P)$ and
$V^{(1)}_{ring}(1+P)$ matrix elements before regularization. The
dashed(red), dash-dotted(green) and dotted(blue) curve represents
the $V^{(1)}_{2\pi}(1+P)$ and $V^{(1)}_{ring}(1+P)$ matrix elements
regularized as in~(\ref{eq.reg}) with $\Lambda=450,550$ and $600$
MeV, respectively. } \label{fig3}
\end{figure}

The examples of the regularized $V^{(1)}(1+P)$  matrix elements are
compared to the nonregularized ones in Fig.~\ref{fig3}. This is done
separately for $V^{(1)}_{2\pi}(1+P)$ and $V^{(1)}_{ring}(1+P)$
contributions. The momenta are $p'=q'=q=0.132$ fm$^{-1}$ (upper row)
and $p'=0.268$ fm$^{-1}$, $q'=2.842$ fm$^{-1}$ and $q=0.132$
fm$^{-1}$ (lower row) and we show only the ($\alpha'=\alpha=1$) channel
combination. In the upper row, where the momenta $p'$, $q'$ and $q$ are
small, all regulator functions are close to $1$ for small values of $p$. For
momenta $p>1$ fm$^{-1}$, the different $\Lambda$ values lead to
different slopes of matrix elements. The lowest value of the parameter
$\Lambda=450$ MeV forces the fastest decreasing of $V^{(1)}_i(1+P)$
matrix elements. In such a case, the short-range part of the interaction
is suppressed. On the contrary, the highest value $\Lambda=600$ MeV
allows for larger contributions of short-range interactions. In the lower row,
where the momenta $p'$ and $q'$ are bigger, the effects of the 
regularization are seen already at the low values of $p$. For $p=0.001$
fm$^{-1}$ the regularization factor $f(p',q') f(p,q)$ changes
from 0.194 for $\Lambda=450$ MeV to 0.747 for $\Lambda=600$ MeV.
This strong cutoff dependence is expected to be largely compensated 
by an appropriate ``running'' of the LECs $c_D$ and $c_E$ when
calculating low-energy observables.  

\section{Determination of the LECs $\fet{c_D}$ and $\fet{c_E}$ at \NNNLO.}
\label{section4}

Once the new terms are added to the 3NF,  the procedure of refitting
of the LECs $c_D$ and $c_E$ has to be repeated. We follow
Ref.\cite{3NF_in_N2LO} and use the triton binding energy $E^{^3H}$
and the neutron-deuteron doublet scattering length $^2a_{nd}$ as two
observables from which $c_D$ and $c_E$ can be obtained. The
up-to-date experimental values are $E^{^3H}=8.481821(5)$
MeV~\cite{exp3H} and $^2a_{nd}=0.645(7)$ fm ~\cite{exp2and}.

Our procedure to fix the values of LECs can be divided into two
steps. First, the dependence of $E^{^3H}$ on $c_E$ for
a given value of $c_D$ is determined. The requirement to reproduce the
experimental value of the triton binding energy yields a set of
combinations $c_D$ and $c_E$. This set is then
used in the calculations of $^2a_{nd}$ what allows us to find which pair of
$c_D$ and $c_E$ describes both observables simultaneously. Such a
procedure has to be repeated for all $\Lambda$-values used in the
regularization. 
The same values of the cutoff $\Lambda$ are used  
to suppress high momenta in the NN potential in order to ensure  the
convergence of the integral in the Lippmann-Schwinger equation.
The
chiral NN potential depends, in addition, on another cut-off parameter
$\tilde{\Lambda}$ emerging from the SFR of the two-pion exchange
potential. We follow Ref.~\cite{Epelbaum} and use five combinations
of ($\Lambda, \tilde{\Lambda}$) shown in Tab.~\ref{tab1}.

We compute the $^3$H wave function using the method 
described in \cite{Nogga.1997}. Here we mention only that the full
triton wave function $\Psi = ( 1 + P ) \psi $ is given by its
Faddeev component $\psi$ being the solution of the Faddeev equation 
\begin{equation}
\psi = G_0 t P \psi + (1+ G_0 t) G_0 V^{(1)} (1+P) \psi \;.
\label{eq.bs}
\end{equation}
Here, $G_0$ is the free 3N propagator, $P$ is the same permutation
operator as defined above and $t$ is the two-body $t$-operator
generated from a given NN potential through the Lippmann-Schwinger
equation.

We use the 3N states $\vert p,q,\alpha
\rangle$ defined on the grids of 68 $p$-points and 48 $q$-points in
intervals $p \in (0,\, 15)$ fm$^{-1}$ and $q\in (0,\, 10)$ fm$^{-1}$,
respectively. We take into account all states up to the two-body total
angular momentum $j=5$ for the NN potential, and all states up to
$j=3$ for the 3N interaction.

\begin{figure}[t]\centering
\epsfig{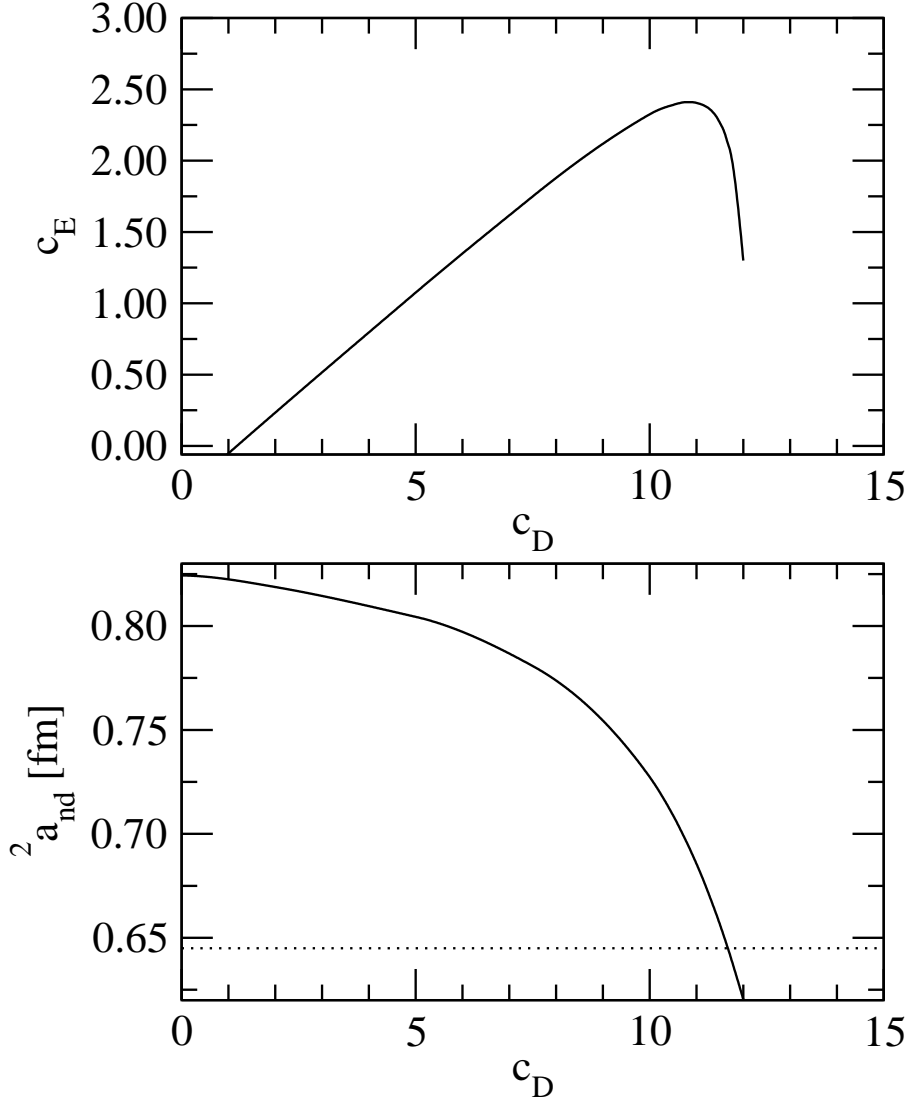}
\caption{(Color online) The intermediate results obtained during
the fitting procedure for the $c_D$ and $c_E$ \NNNLO  LECs for the 3rd
cut-off combination from Tab.~\ref{tab1}: the values of the $c_D$ and
$c_E$ LECs which give the experimental $E^{^3H}$ (top) and the
dependence of $^2a_{nd}$ on $c_D$ (bottom). The dotted line shows
the experimental value $^2a_{nd}$=0.645(7) fm~\cite{exp2and}. }
\label{fig4}
\end{figure}

We solve Eq.(\ref{eq.bs}) and find pairs of the LECs $c_D$ and $c_E$
which reproduce the experimental value of $E^{^3H}$. It is exemplified
in the top panel of Fig.~\ref{fig4} for the 3rd cut-off combination
from Tab.~\ref{tab1}. The dependence is smooth and for 
some values of $c_E$ there are two possible values of $c_D$.

In the second step of the fitting procedure the doublet scattering length $^2a_{nd}$ is
calculated for the ($c_D$, $c_E$) pairs, which reproduce the correct
value of $E^{^3H}$. To
this aim we first solve the Faddeev equation for the auxiliary
amplitude $T$ at zero incoming nucleon energy, 
\begin{equation}
T \ = \ t \, P \, \phi \
+ \ ( 1 + t G_0 ) \, V^{(1)} \, ( 1 + P ) \, \phi \
+ \   \ t \, P \, G_0 \, T \
+ \ ( 1 + t G_0 ) \, V^{(1)} \, ( 1 + P ) \, G_0 \, T \;,
\label{eqT}
\end{equation}
where the initial channel state $\phi$ occurring in the driving
terms is composed of the deuteron and a plane wave state of the
projectile nucleon. The amplitude for the elastic nucleon-deuteron
scattering is then given by
\begin{equation}
U \ = \  P G_0^{-1} \ + \ P T \ + \ V^{(1)} \, ( 1 + P ) \, \phi \
+ \ V^{(1)} \, ( 1 + P ) \, G_0 \, T\;.
 \label{eqU}
\end{equation}
We refer to~\cite{physrep_96, HuberAPP} for a general overview on 3N
scattering and for more details on the practical implementation of the
Faddeev equations.  The expression
for $^2a_{nd}$ in our basis and further technical details can be found
in~\cite{Witala2and}.
In this second step of the fitting procedure we use grids of 32 $p$
points in the range  $p \in
(0,25)$ fm$^{-1}$ and 31 $q$ points in the range $q\in (0,15)$
fm$^{-1}$. Similarly to the triton calculations, the NN (3N) potential
acts in all states up to $j=5(3)$. Our calculations are accurate up to 2 keV 
for the binding energy and up to 0.005 fm 
for the scattering length. We expect that, even in \NNNLO, the chiral 
expansion of the nuclear forces induces uncertainties that are larger 
than these estimates. Therefore, the numerical calculations are 
sufficiently accurate to perform sensible fits of $c_D$ and $c_E$ at \NNNLO.

The final values of $c_D$ and $c_E$ LECs, which 
reproduce the experimental values of
$E^{^3H}$ and $^2a_{nd}$ are given in Tab.~\ref{tab1}.
For all combinations of
cut-off parameters, the LEC $c_D$ remains positive with the value around $10$. It
weakly depends on the value of the cutoff $\Lambda$ and becomes
larger with increasing $\Lambda$. The second LEC $c_E$ changes in a
more complicated way. 
While again it is smallest in magnitude for the smallest value of 
$\Lambda$, its biggest value is for the medium $\Lambda$=550 MeV
and then decreases while moving to $\Lambda$=600 MeV. Note, that
$c_E$ changes sign so the $V^{(1)}_{e-term}$ interaction changes
from attractive to repulsive. We also stress that while the value
of the LEC $c_D$ appears to be rather large, the expectation value of
the one-pion-exchange-contact part of the 3NF is of a 
natural size.  It remains to be seen whether the complete calculation
including the remainig 3NF contributions at N$^3$LO will lead to more
natural values of the LEC $c_D$.

\begin{table}
\begin{tabular}{|c|c|c|c|}
\hline
\;\; cut-off \;\;& ($\Lambda, \tilde{\Lambda})$   & $c_D$ & $c_E$ \\
\hline
1 & \;(450,500)\; & \;\;10.78\;\; & \;\;-0.172\;\; \\
2 & \;(600,500)\; & 12.00  & 1.254  \\
3 & \;(550,600)\; & 11.67 & 2.120   \\
4 & \;(450,700)\; & 7.21  & -0.748 \\
5 & \;(600,700)\; & 14.07 & 1.704 \\
\hline
\end{tabular}
\caption{The values of $c_D$ and $c_E$ LECs for the different
parametrizations of chiral \NNNLO  potential.} \label{tab1}
\end{table}

Finally, we would like to emphasize that the values of the LECs 
which are \emph{bare} parameters must be refitted at each  
order in the chiral expansion (and, of course, for each cutoff
combination). 
This is in contrast with  chiral perturbation theory calculations 
in the Goldstone-boson and
single-nucleon sectors where the scattering amplitude is usually
expressed in terms of \emph{renormalized} LECs. Indeed,  using 
the values of $c_D$ and $c_E$ determined at  \NNLO in the \NNNLO
calculation would generally result in a poor description of low-energy
observables. For
example, for the 3rd cutoff combination, the 3NF at  \NNNLO  furnished
with the \NNLO values 
$c_D=-0.45$ and $c_E=-0.798$~\cite{Epelbaum} yields 
$E^{^3H}=-8.197$ MeV and $^2a_{nd}=1.004$ fm which are far from
the experimental values.
Similarly, while the combination $c_D=1.5744$ and $c_E=-17.8$ 
allows
to reproduce the triton binding energy and the doublet $nd$ scattering
length for \NNNLO NN force accompanied with the \NNLO 3NF, it produces 
$E^{^3H}=-7.542$ MeV and $^2a_{nd}=1.4354$ fm when the \NNNLO 3NF is
used. 
Note that a big value of $c_E$ obtained in this case seems to 
violate naturalness.
These results demonstrate clearly that fitting to the data has
to be made consistently within the given
order of the chiral expansion. 
Therefore, also the results of our test fit here, which does not include 
the full 3NF at \NNNLO, has to be taken with care. However, the fit 
results allows us to study the properties of $^3$H in the next section. 

\section{The properties of $\fet{^3}$H}
\label{section5}

Once the values of $c_D$ and $c_E$ low energy parameters are
established, one can explore the properties of the $^3$H wave
function. We begin with the expectation values of the kinetic
energy $\langle H_0 \rangle$, the NN potential energy $\langle
V_{NN} \rangle$ and the 3N potential energy $\langle V_{3N}
\rangle$ which are listed in Tab.~\ref{tab2}.
\begin{table}
\begin{tabular}{|c|c|c|c|}
\hline
cut-off     & $\langle H_0 \rangle$ [MeV] & $\langle V_{NN} \rangle$ [MeV] & $\langle V_{3N} \rangle$ [MeV] \\
\hline
1      &  35.972 & -43.459 & -0.994 \\
2      &  54.708 & -61.515 & -1.673 \\
3      &  48.088 & -55.187 & -1.381 \\
4      &  33.232 & -41.050 & -0.663 \\
5      &  53.504 & -60.278 & -1.706 \\
\hline
\end{tabular}
\caption{Expectation values $\langle H_0 \rangle$,
$\langle V_{NN} \rangle$ and $\langle V_{3N} \rangle$ in the triton for different
parametrizations of the chiral \NNNLO  potential as discussed in the text.} \label{tab2}
\end{table}
The expectation values clearly depend on the
cut-off parameters $\Lambda$ and $\tilde{\Lambda}$ as they should. Not
surprisingly, the expectation values of both the NN potential and the
3NF are smallest for the softest cutoff $\Lambda=450$ MeV. 
Higher $\Lambda$-values lead to stronger 3NF contributions 
to the $E^{^3H}$ which for $\Lambda=600$ MeV reaches about 
 $\approx 20\%$. For all cutoff values, one clearly observes the
 dominance of the NN forces, $\langle V_{3N} \rangle/\langle V_{NN} \rangle =
 1.5 \ldots 3\%$, in agreement with the expectations based on the
 chiral power counting. 
Note that the interplay of \NNLO counter terms and \NNNLO structures of the 3NF does 
not allow to use these expectation values for an assessment  of the contributions 
of \NNNLO 3NFs. Similarly as in the NN interaction, one observes strong cancelations 
of \NNNLO contributions also with \NNLO contact interactions. Such a comparison 
only makes sense for renormalized quanties, which we are not able to identify here.   
One also observes that the dependence of the
 expectation values on the SFR cutoff $\tilde{\Lambda}$ is less
 pronounced as the $\Lambda$-dependence.  For
example for cutoff combinations 1 and 4, which differ only in the
choice of
$\tilde{\Lambda}$, the $\langle H_0 \rangle$ and $\langle V_{NN}
\rangle$ differ by about $3$ MeV. On the other hand, the differences
reach almost $20$ MeV for the cutoff combinations 1 and 2 which have
the same SFR cutoff $\tilde \Lambda$ but different values of
$\Lambda$.    This holds true also  for $\langle V_{3N} \rangle$.

A more detailed information about the 3NF triton expectation values is given
in Tab.~\ref{tab3}. The expectation value of the 3NF $\langle V_{3N}
\rangle$ is split into the individual contributions from various
topologies. The
expectation value of the two-pion-exchange potential $\langle
V_{2\pi} \rangle$ shows a smooth dependence with $\Lambda$. For the
softest cutoff $\Lambda=450$ MeV, the two-pion exchange 3NF turns out
to be most attractive providing more than $0.5$ MeV to the triton
binding energy.  With increasing $\Lambda$, the  contribution of
$V_{2\pi}$ becomes weaker. For the cutoff combination 2, $\Lambda =
600$ MeV, $\tilde \Lambda = 500$ MeV, the additional binding due to
the two-pion exchange 3NF only amounts to about $240$ keV.
Interestingly, most of the attraction necessary to reproduce the
triton binding energy is produced in this case by the ring topology,
which is found to be attractive for all cutoff combinations. 
Contrary to the longest-range two-pion exchange topology, the
contributions of the ring diagrams are enhanced for the largest
value of the cutoff  $\Lambda = 600$ MeV. Qualitatively,
this behavior might be expected given the fact that the large values
of $\Lambda$ probe the shorter-range part of $V_{ring}$ which is of
the van der-Waals type, i.e.~the matrix elements grow rapidly with
decreasing relative distances between the nucleons. 
The cutoff dependence of this contribution is very large. This explicitly 
shows the dependence on the short distance pieces making it impossible 
to estimate the impact of this topology on low energy observables 
based on our results. 
 The expectation
value of the two-pion-one-pion-exchange topology $\langle V_{2\pi-1\pi}
\rangle$ also strongly depends on  $\Lambda$. It  changes
sign from positive for  $\Lambda=450$ MeV to negative at
$\Lambda=600$ MeV. The dependence on the SFR cutoff is fairly weak. 
Note however that a stronger dependence might be induced once the SFR regularized $A(q)$ 
has been taken into account. 
The $\langle V_{d-term} \rangle$ shows the most
complicated behaviour. It achieves the lowest value for the
intermediate value of $\Lambda$ (cutoff combination 3) and, for 
$\Lambda=600$ MeV, shows a strong dependence on the SFR cutoff $\tilde
\Lambda$.  In particular, for the smaller value 
$\tilde{\Lambda}= 500$ MeV, this contribution to the 3NF becomes
repulsive and relatively big, while for $\tilde{\Lambda}=700$ MeV the
expectation value remains
negative. Finally, the $\langle V_{e-term} \rangle$ expectation
value changes smoothly with $\Lambda$. It also changes its sign from 
positive at the lowest $\Lambda$ to negative at $\Lambda=600$
MeV.
\begin{table}
\begin{tabular}{|c|c|c|c|c|c|}
\hline
cut-off  & $\langle V_{2\pi} \rangle$ [MeV] & $\langle V_{2\pi-1\pi} \rangle$ [MeV] & $\langle V_{ring} \rangle$ [MeV] & $\langle V_{d-term} \rangle$ [MeV] & $\langle V_{e-term} \rangle$ [MeV] \\
\hline
1      & -0.639 &  0.458 & -0.147 & -0.693 & 0.027  \\
2      & -0.241 & -0.580 & -1.114 &  0.694 & -0.432 \\
3      & -0.473 &  0.107 & -0.191 & -0.708 & -0.116 \\
4      & -0.771 &  0.539 & -0.452 & -0.259 & 0.281  \\
5      & -0.377 & -0.275 & -0.622 & -0.119 & -0.313 \\
\hline
\end{tabular}
\caption{The expectation values for the different parts of the 3N potential and
for the different parametrizations of the chiral \NNNLO  potential.}
\label{tab3}
\end{table}
Last but not least, we emphasize that the expectation values discussed
above as well as the separation of the potential energy into the
contributions due to the NN potential and 3NF do \emph{not} correspond
to observable quantities and are expected to show a strong cutoff
dependence. Notice further that expectation values of the various 3NF
contributions are, strictly speaking, bare quantities. Comparing their
size with the one of the N$^2$LO terms does, therefore, not allow to
draw conclusions about the convergence of the chiral
expansion~\cite{footnote}.
 It is comforting to see that all expectation
values turn out to be of a reasonable size.   

\begin{figure}[t]\centering
\epsfig{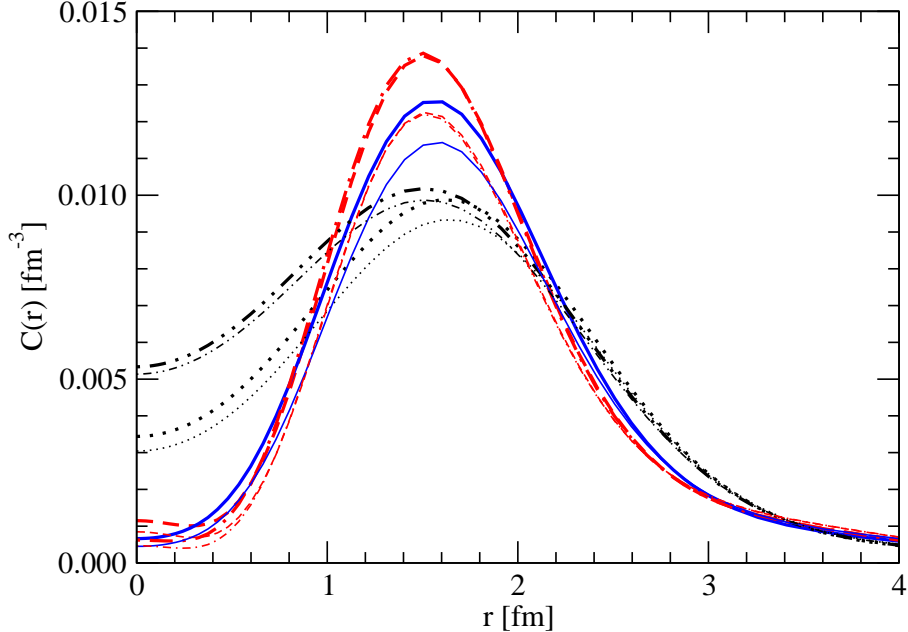}
\caption{(Color online) The two-body correlation function
for the triton for different $(\Lambda, \tilde{\Lambda})$ pairs
given in Tab.~\ref{tab1}. The thin dotted (black), dashed (red), solid (blue),
dash-double-dotted (black) and dash-dotted (red) lines correspond to predictions based only on
NN interaction with cut-off numbers from 1 to 5, respectively.
The thick lines represent predictions with the same regularization parameters
but based on NN and 3N forces.}
\label{fig5}
\end{figure}

We now turn to the two-nucleon correlation function of $^3$H which is defined
as~\cite{Nogga.1997}
\begin{equation}
C(r) \equiv \frac13 \frac{1}{4\pi} \int d\hat{r} \langle \Psi \mid \sum_{i<j} \delta(\vec{r}-\vec{r_{ij}}) \mid \Psi \rangle\,.
\end{equation}
Here, $\vec{r_{ij}}$ is the relative distance corresponding to
the Jacobi momentum $\vec{p}$. In Fig.~\ref{fig5}
the correlation function is shown for the same combinations of the 
regularization parameters as in 
Tab.\ref{tab1}. Thin lines represent predictions based on NN
interactions only while thick ones show the predictions based on NN+3N
forces. As expected, the softest cutoff value $\Lambda=450$ MeV yields
a more flat correlation function with the less amount of the
short-range correlations. The higher
$\Lambda$ values prefer distributions concentrated around the maximum at
$r \approx 1.5$ fm. The effects of the 3NFs are small for the lowest
$\Lambda$ but increase with increasing $\Lambda$. For the lowest
$\Lambda$ there is also a strong dependence of the correlation
function on the SFR parameter $\tilde{\Lambda}$.

\section{Summary and outlook}
\label{conclusion}

In this paper we, for the first time, included 
the long-range \NNNLO corrections to the 3NF in the Faddeev
calculations. These novel two-pion-exchange, two-pion-one-pion and
ring interactions supplemented with the one-pion-exchange and contact
terms emerging at \NNLO represent, presently, the most advanced chiral 3NF.
We use this force in the triton and the doublet neutron-deuteron
scattering length Faddeev calculations to fix the two  
low-energy parameters $c_D$ and $c_E$ for different sets of
regularization parameters, which cut off large-momentum or,
equivalently, short-range components in the few-nucleon states. 
While the value of the LEC $c_D$ remains fairly stable,  $c_E$
features a stronger
sensitivity to regularization parameters. 
It will be interesting whether such a behavior is also seen 
for fits involving the complete \NNNLO 3NF. 
We also studied the
individual contributions of the various topologies to the
triton binding energy.  The expectation
values of the two-pion-one-pion and ring terms turn out to be 
smaller than the ones of the
dominant two-pion-exchange 3NF for softer values of the regulator. 
Generally, all expectation values are found to be sizable.  As expected,
we observe a strong sensitivity of the 
expectation values  to the regularization parameters.
We also looked at the impact of the used 3NF on the two-nucleon
correlation function in the triton. 

While our work does not yet correspond to a complete N$^3$LO
analysis due to the shorter-range contributions and
relativistic corrections to the 3NF which are not yet available and 
are still missing in our calculations, it does represent a very important step 
in this direction and provides a proof-of-principle that the very
complex operator structure of the 3NF at \NNNLO can be successfully
implemented in few-body calculations. In the future, this study should
be extended to explore effects of the novel terms in the 3NF in few-nucleon
scattering. This work is in progress. 
To complete the analysis of the 3NF at \NNNLO  the inclusion of full
structure of shorter-range $V^{(1)}_{1\pi-cont}$,
$V^{(1)}_{2\pi-cont}$ as well as  $V^{(1)}_{1/m}$ terms should be
pursued. The numerical implementation of the new terms can be 
straightforwardly performed using the newly developed aPWD scheme,
which is successfully tested for the
long-range terms in the present study. 
Finally, it should be emphasized that the present work also opens the
way for applying the novel chiral nuclear forces in many-body
calculations, see e.g.~\cite{Navratil:2009ut,Roth:2011ar,Lesinski:2011rn} for some exciting recent developments
 along these lines based on \NNLO 3NFs.

\section*{Acknowledgments}

This work was supported by
the Polish Ministry of Science and Higher Education
under Grant No. N N202 077435.
It was also partially supported by the Helmholtz
Association  (grants VH-NG-222 and
VH-VI-231), by the European Community-Research Infrastructure
Integrating Activity
``Study of Strongly Interacting Matter'' (acronym HadronPhysics2,
Grant Agreement n. 227431)
under the Seventh Framework Programme of EU and 
the European Research Council (ERC-2010-StG 259218 NuclearEFT). 
The numerical
calculations have been performed on the
 supercomputer cluster of the JSC, J\"ulich, Germany
 and ACK-Cyfronet, Krak\'ow, Poland.




\begin{thebibliography}{99}
\bibitem{footnote}
In fact, $r$-space potentials generated by the
  long-range N$^3$LO contributions are much weaker than the
  ones emerging from the leading two-pion exchange terms at relative
  distances of the order of the inverse pion mass.
\end{thebibliography}
\end{document}